\documentstyle [12pt,axodraw,psfig] {article}

\def\gsim{\:\raisebox{-0.5ex}{$\stackrel{\textstyle>}{\sim}$}\:}

\catcode`@=12
\begin{document}
\thispagestyle{empty}
\begin{flushright} UCRHEP-T224\\TIFR/TH/98-20\\May 1998\
\end{flushright}
\vspace{0.5in}
\begin{center}
{\Large	\bf Phenomenology of the $B - 3 L_\tau$ Gauge Boson\\}
\vspace{1.5in}
{\bf Ernest Ma\\}
\vspace{0.1in}
{\sl Department of Physics, University of California\\}
{\sl Riverside, California 92521, USA\\}
\vspace{0.3in}
{\bf D. P. Roy\\}
\vspace{0.1in}
{\sl Tata Institute of Fundamental Research, Mumbai 400 005, India\\}
\vspace{1.5in}
\end{center}
\begin{abstract}\
Assuming the existence of a gauge boson $X$ which couples to $B - 3 L_\tau$, 
we discuss the present experimental constraints on $g_X$ and $m_X$ from 
$Z \rightarrow l^+ l^-$ and $Z \rightarrow \bar f f X~(f = q, \nu_\tau, 
\tau)$.  We 
also discuss the discovery potential of $X$ at hadron colliders through its 
decay into $\tau^+ \tau^-$ pairs.  In the scenario where all three charged 
leptons (and their neutrinos) mix, lepton flavor nonconservation through $X$ 
becomes possible and provides another experimental probe into this hypothesis.
\end{abstract}

\newpage
\baselineskip 24pt

\section{Introduction}

In the minimal standard model of quarks and leptons, neutrinos ($\nu_e$, 
$\nu_\mu$, $\nu_\tau$) appear only 
as members of left-handed doublets and there is a single Higgs scalar 
doublet.  Hence neutrinos are massless and each lepton number ($L_e$, 
$L_\mu$, $L_\tau$) is separately conserved at the classical level as is 
baryon number ($B$).  However, only the linear combination $B - L_e - L_\mu 
- L_\tau$ remains conserved at the quantum level\cite{1}, whereas the 
corresponding U(1) is still anomalous and cannot be gauged.  Actually, one 
of the three lepton number differences ($L_e - L_\mu$, $L_e - L_\tau$, 
$L_\mu - L_\tau$) is anomaly-free and could be gauged\cite{2}.  On the other 
hand, this particular extension of the standard model would not shed much 
light on the question of neutrino mass.  After all, there is now a host of 
experimental evidence for neutrino oscillations and that can be explained 
most naturally if neutrinos have masses and mix with one another.  The 
canonical way of doing this is to add three right-handed neutrino singlets 
with large Majorana masses so that $\nu_e$, $\nu_\mu$, and $\nu_\tau$ all 
obtain small seesaw masses.  Such an extension of the standard 
model allows $B - L$ to be gauged\cite{3}. [Since the three lepton families 
now mix, it makes sense to consider only one lepton number, {\it i.e.} 
$L = L_e + L_\mu + L_\tau$.]

Suppose we add only \underline {one} right-handed neutrino singlet and pair 
it with $\nu_\tau$.  Then the symmetry $B - 3 L_\tau$ can be gauged\cite{4}. 
Just as $B - L$ may originate\cite{5} from $SU(4) \times SU(2)_L \times 
SU(2)_R$, the breaking of $SU(10) \times SU(2)_L \times U(1)_{Y'}$ to the 
standard gauge group by way of $SU(9)$ leads naturally\cite{4} to $B - 
3 L_\tau$.  This recent discovery opens up a possible rich phenomenology 
associated with the $B - 3 L_\tau$ gauge boson which we call $X$.  The key 
observation is that $X$ is not constrained by present experimental data 
to be very heavy because it does not couple to leptons of the first and 
second families.

In Sec.~2 we describe the $B - 3 L_\tau$ model and show how all anomalies 
are canceled.  In Sec.~3 we determine the present experimental constraints 
on the mass and coupling of $X$.  They come chiefly from the nonobservation 
of $X$ in the decay of $Z$ and the agreement with $e - \mu - \tau$ 
universality in $Z$ decays.  In Sec.~4 we consider the production of $X$ 
at hadron colliders and the prospect for its detection at the Tevatron 
and at the LHC (Large Hadron Collider).  In Sec.~5 we study how $\nu_e$ 
and $\nu_\mu$ may acquire radiative masses and mix with $\nu_\tau$ which 
has a tree-level seesaw mass.  We present two scenarios, one with 
lepton-flavor-changing couplings for $X$ and one without.  In Sec.~6 we 
discuss the possible manifestations of lepton flavor nonconservation in 
the first scenario.  We also explain how the one-loop quark flavor 
nonconservation is naturally suppressed in this model.  Finally in Sec.~7 
we have some concluding remarks.

\section{Structure of the $B - 3 L_\tau$ Gauge Model}

Consider the gauge group $SU(3)_C \times SU(2)_L \times U(1)_Y \times U(1)_X$, 
where the extra U(1) refers to the $B - 3 L_\tau$ symmetry.  The quarks and 
leptons transform thus as follows.
\begin{equation}
\left( \begin{array} {c} u_i \\ d_i \end{array} \right)_L \sim (3,2,1/6;1/3), 
~~~ u_{iR} \sim (3,1,2/3;1/3), ~~~ d_{iR} \sim (3,1,-1/3;1/3);
\end{equation}
\begin{equation}
\left( \begin{array} {c} \nu_e \\ e \end{array} \right)_L, \left( 
\begin{array} {c} \nu_\mu \\ \mu \end{array} \right)_L \sim (1,2,-1/2;0), 
~~~ e_R, \mu_R \sim (1,1,-1;0);
\end{equation}
\begin{equation}
\left( \begin{array} {c} \nu_\tau \\ \tau \end{array} \right)_L \sim 
(1,2,-1/2;-3), ~~~ \tau_R \sim (1,1,-1;-3), ~~~ \nu_{\tau R} \sim 
(1,1,0;-3).
\end{equation}
In the above, only one right-handed neutrino singlet, {\it i.e.} 
$\nu_{\tau R}$, has been added to the minimal standard model.  Since the 
number of $SU(2)_L$ doublets remains even (it is in fact unchanged), the 
global SU(2) chiral gauge anomaly\cite{6} is absent.  Since the quarks 
and leptons are chosen to transform vectorially under the new $U(1)_X$, 
the mixed gravitational-gauge anomaly\cite{7} is also absent.  The various 
axial-vector anomalies\cite{8} are canceled as well.  The $[SU(3)]^2 U(1)_X$ 
and $[U(1)_X]^3$ anomalies are automatically zero because of the vectorial 
nature of $SU(3)$ and $U(1)_X$.  The remaining conditons are satisfied as 
follows.
\begin{eqnarray}
[SU(2)]^2 U(1)_X &:& (3)(3)(1/3) + (-3) = 0; \\ ~ 
[U(1)_X]^2 U(1)_Y &:& (3)(3)(1/3)^2[2(1/6)-(2/3)-(-1/3)] \nonumber \\ && + 
(-3)^2[2(-1/2)-(-1)] = 0; \\  ~
[U(1)_Y]^2 U(1)_X &:& (3)(3)[2(1/6)^2 - (2/3)^2 - (-1/3)^2](1/3) \nonumber \\ 
&& + [2(-1/2)^2 - (-1)^2](-3) = 0.
\end{eqnarray}

The minimal scalar content of this model consists of just the usual doublet 
\begin{equation}
\left( \begin{array} {c} \phi^+ \\ \phi^0 \end{array} \right) \sim 
(1,2,1/2;0)
\end{equation}
and a neutral singlet
\begin{equation}
\chi^0 \sim (1,1,0;6)
\end{equation}
which couples to $\nu_{\tau R} \nu_{\tau R}$.  As the former acquires a 
nonzero vacuum expectation value, the electroweak gauge symmetry $SU(2)_L 
\times U(1)_L$ breaks down to $U(1)_{em}$, whereas $\langle \chi^0 \rangle 
\neq 0$ breaks $U(1)_X$.  The resulting theory allows $\nu_{\tau L}$ to 
obtain a seesaw mass and retains $B$ as an additively conserved quantum 
number and $L_\tau$ as a multiplicatively conserved quantum number.  The 
two other neutrinos, {\it i.e.} $\nu_e$ and $\nu_\mu$, are massless in this 
minimal scenario and cannot mix with $\nu_\tau$.  In fact, $L_e$ and $L_\mu$ 
are still separately conserved, and $L_e - L_\mu$ can still be gauged at 
this point.  To obtain a phenomenologically interesting neutrino sector, 
{\it i.e.} to accommodate present neutrino-oscillation data, we will consider 
later two scenarios for extending the scalar sector to allow $\nu_e$ and 
$\nu_\mu$ to acquire radiative masses and to mix with $\nu_\tau$.

\section{Constraints on $X$ from $Z$ Decay}

Since the $B - 3 L_\tau$ gauge boson $X$ does not couple to $e$ or $\mu$ or 
their corresponding neutrinos, 
there is no direct phenomenological constraint from the best known 
high-energy physics experiments, such as $e^+ e^-$ annihilation, 
deep-inelastic scattering of $e$ or $\mu$ or $\nu_\mu$ on nuclei, or 
the observation of $e^+ e^-$ or $\mu^+ \mu^-$ pairs in hadronic collisions. 
Although $X$ does contribute to purely hadronic interactions, its presence 
is effectively masked by the enormous background due to quantum 
chromodynamics (QCD).  However, unlike the case of a gauge boson coupled 
only to baryon number\cite{9}, $X$ also couples to $L_\tau$.  Hence $X$ 
may decay into $\tau^+ \tau^-$ or $\bar \nu_\tau \nu_\tau$ and be detected 
that way if it is produced. 

The mass and coupling of $X$ are constrained by present experimental
data (from LEP, mainly) in two important ways.  The first is direct
production through $Z$ decay:  
\begin{equation}
Z \rightarrow (\bar q q, \tau^+ \tau^-, \bar \nu_\tau \nu_\tau) + X, ~~ 
{\rm then} ~ X \rightarrow (\bar q q, \tau^+ \tau^-, \bar \nu_\tau \nu_\tau).
\end{equation}
This applies of course only to $m_X < M_Z$.  The second is through its 
radiative contribution to $Z \rightarrow (\tau^+ \tau^-, \bar \nu_\tau 
\nu_\tau)$ which breaks $e - \mu - \tau$ universality.

Because of the $B - 3 L_\tau$ gauge symmetry, the branching fractions of $X$ 
to $\tau^+ \tau^-$ and $\bar \nu_\tau \nu_\tau$ are very substantial. 
Assuming that $\nu_{\tau R}$ and the $t$ quark are too heavy to be decay 
products of $X$, and using the parton model as a crude approximation, the 
branching fractions of $X$ are
\begin{eqnarray}
B(X \to \tau^+ \tau^-) &=& 54/91 = 0.59, \\ 
B(X \to \bar \nu_\tau \nu_\tau) &=& 27/91 = 0.30, \\ 
B(X \to \bar q q) &=& 10/91 = 0.11.
\end{eqnarray}

Consider $Z \to \bar f f X$, where $f = \tau, \nu_\tau, q$.  In the 
center-of-momentum frame, let $E_1$ and $E_2$ be the energies of $f$ and 
$\bar f$, and $\theta$ the angle between their directions.  Then the 
square of the amplitude averaged over the polarizations of $Z$ is easily 
calculated to be
\begin{eqnarray}
|\overline {\cal M}|^2 &=& (B - 3 L_\tau)^2 g_X^2 \left( {g_L^2 + g_R^2 \over 
2} \right) (16 E_1 E_2) \left\{ (1 + \cos \theta) \left[ {1 \over (M_Z - 
2 E_1)^2} + { 1 \over (M_Z - 2 E_2)^2} \right] \right. \nonumber \\ 
&& + \left. {4(1 - \cos \theta) \over (M_Z - 2 E_1) (M_Z - 2 E_2)} 
\left[ 1 - {E_1 + E_2 \over M_Z} + {E_1 E_2 (1 - \cos \theta) \over M_Z^2} 
\right] \right\},
\end{eqnarray}
where $g_L = (g/\cos \theta_W)(I_{3L} - \sin^2 \theta_W Q)$ and $g_R = 
(g/\cos \theta_W)(- \sin^2 \theta_W Q)$ 
are the standard-model couplings of $f$ and $\bar f$ to $Z$.

Of the 9 possible final-state combinations of $Z$ decaying into $X$, two are 
very amenable to experimental detection, {\it i.e.}
\begin{eqnarray}
&& Z \to \bar q q + X, ~~ {\rm then} ~X \to \bar \nu_\tau \nu_\tau, \\ 
&& Z \to \bar \nu_\tau \nu_\tau + X, ~~ {\rm then} ~X \to \bar q q.
\end{eqnarray}
Both result in 2 jets plus missing energy.  This channel has been
widely investigated for the Higgs-boson search at LEP-I.  The decay
process (14) resembles the Higgs signal for its invisible decay into
majorons \cite{10,11}, while (15) resembles the signal for its SM (Standard 
Model) decay.
The total ALEPH data from LEP-I, corresponding to 4.5 million hadronic
$Z$ events, show no events in the 2 jets plus missing energy channel
after the selection cuts \cite{12}.  We have analysed these data in terms of
the decay processes (14,15) using a parton level Monte Carlo program.
The program was earlier shown to reproduce the effeciencies of these
selection cuts very well in the context of the SM Higgs signal [10];
and it is expected to work equally well here.  In the absence of any
candidate events, the 95\% CL (confidence-level) limit on the signal 
corresponds to 3
events after the selection cuts.  The resulting upper limit on $g_X$
is shown as a function of $m_X$ in Fig. 1.  One gets a stringent limit
on $g_X$ $(< 0.1)$ for $m_X < 50 \ {\rm GeV}$, where the signal has a
reasonable efficiency of $\sim 40\%$.  But it deteriorates rapidly for
$m_X \geq 70$ GeV, where the efficiency for the dominant process (15) goes
down due to a low missing energy. 

Three other final states are also important, {\it i.e.}
\begin{eqnarray}
&& Z \to \tau^+ \tau^- + X, ~~ {\rm then} ~X \to \bar \nu_\tau \nu_\tau, \\ 
&& Z \to \bar \nu_\tau \nu_\tau + X, ~~ {\rm then} ~X \to \tau^+ \tau^-, \\ 
&& Z \to \tau^+ \tau^- + X, ~~ {\rm then} ~X \to \tau^+ \tau^-.
\end{eqnarray}
The first two have two $\tau$'s plus missing energy in the final
state, for which the combined decay rate is about 5 times as large as
that of (14,15).  Unfortunately the bulk of the LEP-I data in this
channel have not been processed, since $H \rightarrow \tau\tau$ is
not an important decay channel for the Higgs mass range of current
interest.  The only useful data we could find come from an early
Higgs search program of ALEPH, corresponding to $< 0.2$ million
hadronic $Z$ events \cite{13}.  The efficiency factor for this channel 
is slightly less than that for (14,15).  Nonetheless one expects a factor 
of $\sim 2$ improvement in the $g_X$ limit if the full ALEPH data in this 
channel are analysed.  The last process (18) corresponds to a
$4\tau$ channel and has twice as large a decay rate as (14,15).  The
detection efficiency for this channel has been estimated to be about
30\% for the SM background, where the extra pair of $\tau$'s come from
a virtual $\gamma/Z$ \cite{14}.  The size of this background for the
total data sample of 4.5 million hadronic $Z$ events is estimated to
be 2-3 events, which can be subtracted out.  Assuming a similar
detection efficiency for the signal process (18), one expects to get at
least as good a limit on $g_X$ from this channel as from (14,15).

The second constraint on $g_X$ and $m_X$ comes from the observed universality 
of $Z \to l^+l^-$ decays.  Since the one-loop radiative correction of the 
$Z \tau^+ \tau^-$ vertex has an extra contribution from $X$, a small 
deviation from $e - \mu - \tau$ universality is expected.  From the precision 
measurements\cite{15} at the $Z$ resonance, {\it i.e.}
\begin{equation}
\Gamma_e = 83.94 \pm 0.14 ~{\rm MeV}, ~~~ \Gamma_\mu = 83.84 \pm 0.20 ~
{\rm MeV}, ~~~ \Gamma_\tau = 83.68 \pm 0.24 ~{\rm MeV},
\end{equation}
and adding 0.19 MeV to $\Gamma_\tau$ to adjust for the kinematical correction 
due to $m_\tau$, we find the deviation of $\Gamma_\tau$ from the average of 
$\Gamma_e$ and $\Gamma_\mu$ to be bounded at 95\% CL as follows:
\begin{equation}
\Delta \Gamma_\tau / \Gamma_{e,\mu} < 0.006.
\end{equation}
Let $\delta \equiv m_X^2/M_Z^2$, then the one-loop radiative correction to 
$Z \to \tau^+ \tau^-$ from $X$ exchange is given by\cite{9}
\begin{equation}
{\Delta \Gamma_\tau \over \Gamma_\tau} = {9 g_X^2 \over 8 \pi^2} 
F_2 (\delta),
\end{equation}
where it is well-known that
\begin{eqnarray}
F_2 (\delta) &=& -2 \left\{ {7 \over 4} + \delta + \left( \delta + {3 \over 2} 
\right) \ln \delta \right. \nonumber \\ && \left. + (1 + \delta)^2 \left[ 
{\rm Li}_2 \left( {\delta \over 1 + 
\delta} \right) + {1 \over 2} \ln^2 \left( {\delta \over 1 + \delta} \right) 
- {\pi^2 \over 6} \right] \right\}.
\end{eqnarray}
In the above, ${\rm Li}_2(x) = -\int_0^x {dt \over t} \ln (1-t)$ is the 
Spence function.  Using the experimental bound of Eq.~(20) we show in 
Fig.~1 the upper limit (dashed line) on $g_X$ as a function of $m_X$.  
Since the function $F_2$ decreases only slowly as $\delta$ increases, we 
find that the upper limit on $g_X$ increases from 0.22 at $m_X = M_Z$ 
to only 0.32 at $m_X = 2 M_Z$.

From the constraint on the invisible width of the $Z$ which measures\cite{15} 
the effective number of neutrinos to be $2.993 \pm 0.011$, we find 
$\Delta \Gamma_{\nu_\tau} / \Gamma_{\nu_\tau} < 0.015$.  Since this 
quantity is also determined by the right-hand side of Eq.~(21), it is 
superceded by the bound of Eq.~(20).

Assuming $g_1 \simeq 0.35$ to represent the typical size of a
$U(1)$ gauge coupling, we see that the universality limit of $g_X$ is
a fairly significant result.  However it does not rule out any range
of $m_X$.  On the other hand, the $Z$-decay limit seems to rule out $m_X
\leq 40 \ {\rm GeV}$, since the corresponding limit on $g_X$ $(\leq
0.05)$ is an order of magnitude smaller than $g_1$. 

\section{Production and Detection of $X$ at Hadron Colliders: Present 
Constraint and Future Prospect}

The large branching fraction of the $X \rightarrow \tau\tau$ decay can
be exploited to search for $X$ in the $\tau\tau$ channel at
hadron colliders.  We shall consider the leptonic decay of one $\tau$
and hadronic decay of the other, resulting in a $l\tau$ final
state.  Recently the CDF collaboration have presented their total
$l\tau$ data from the Tevatron (Run I), corresponding to an
integrated luminosity of $110 \ {\rm pb}^{-1}$ \cite{16}.  The details
of the data along with the selection cuts can be found in the second
paper of \cite{16}.  A large fraction of this $l\tau$ data set contains
a single accompanying jet.  Of the 22 observed events in this data sample, 11
are estimated to come from 
\begin{equation}
Z \rightarrow \tau\tau,
\end{equation}
while most of the rest are estimated as $\tau$ fakes.  We
have estimated the $X$ contribution to this channel using a
parton level Monte Carlo program, which was found to reproduce the
size of the above $Z$ contribution reasonably well.  As in the case of
$Z$, the relevant production processes for $X$ are the NLO (next to 
leading order) Drell-Yan processes,
\begin{equation}
q\bar q \rightarrow gX \ ~{\rm and}~ \ gq (\bar q) \rightarrow q (\bar
q) X.
\end{equation}

With 22 observed events and a background of similar magnitude, the
95\% CL limit on the $X$ boson signal can be estimated to be about 12
events \cite{17}.  The corresponding limit on $g_X$ is shown as the
solid line in Fig. 2.  The plot is shown for $m_X \geq 100 \ {\rm
GeV}$, since for a light $X$ the final lepton from $X \rightarrow
\tau \rightarrow l$ decay becomes too soft to survive the selection
cut.  Thus there is a complementarity between the $X$ search at hadron
colliders and in $Z$-decay at LEP-I.

Note that the present Tevatron limit on $g_X$ is not much better than
the universality limit from LEP-I (Fig. 1).  However, there is scope
for significant improvement of this limit with much larger data samples
expected from Tevatron (Run II) and LHC.  In that case one can
separate the $X$ signal from $Z$ by imposing a $p_T$ cut on $X(Z)$,
which will enable one to reconstruct the momenta of the decay $\tau$
pair and hence the $X(Z)$ mass.  This technique has been widely investigated in the context of
Higgs boson search at hadron colliders in the $\tau\tau$ channel.  We
have explored this quantitatively by imposing a $p^{X(Z)}_T > 50 \
{\rm GeV}$ cut for TeV-II and $p_T^{X(Z)} > 100 \ {\rm GeV}$ for the LHC.
The resulting discovery limits of TeV-II and LHC are shown in Fig. 2.
They correspond to 10 signal events with the expected luminosities of
$2 \ {\rm fb}^{-1}$ at TeV-II and $10 \ {\rm fb}^{-1}$ at LHC.  The
latter corresponds to the low luminosity run of LHC.  Even with this
run it should be possible to probe for the $X$ boson up to $m_X = 500 \
{\rm GeV}$, assuming that $g_X$ is of the same order of magnitude as
$g_1$. The probe can be extended up to $m_X = 1 \ {\rm TeV}$ at the high 
luminosity 
run of LHC, which is expected to deliver an integrated luminosity of 
$100 \ {\rm fb}^{-1}$. 

\section{Radiative Neutrino Masses}

In the presence of the $B -3 L_\tau$ gauge symmetry, only $\nu_\tau$ has 
a right-handed partner and thereby a seesaw mass.  To 
accommodate the present data on neutrino oscillations, we need to allow 
$\nu_e$ and $\nu_\mu$ to be massive, and have them mix with 
each other and $\nu_\tau$.  To this end, we must break the remaining 
leptonic symmetries, {\it i.e.} multiplicative $L_\tau$ as well as additive 
$L_e$ and $L_\mu$.  One possible scenario was already proposed in Ref.~[4]. 
The scalar sector is extended to include a doublet
\begin{equation}
\left( \begin{array} {c} \eta^+ \\ \eta^0 \end{array} \right) \sim 
(1,2,1/2;-3) 
\end{equation}
and a charged singlet
\begin{equation}
\chi^- \sim (1,1,-1;-3).
\end{equation}
The doublet breaks $L_e$, $L_\mu$, and $L_\tau$ separately but an overall 
multiplicative lepton number is preserved.  It also generates flavor-changing 
couplings of $X$ to the charged leptons, details of which will be discussed 
in the next section.  The singlet $\nu_{\tau R}$ is now paired with one 
linear combination of the three left-handed neutrinos.  It appears at first 
sight that there are then two massless neutrinos left.  However, since the 
three lepton numbers are no longer individually conserved, these neutrinos 
necessarily pick up radiative masses.  This generally happens in two loops 
through double $W$ exchange \cite{18}, but the masses so obtained are 
extremely small.  In the present scenario without the $\chi^-$ singlet, 
one of the two massless neutrinos at tree level does pick up a radiative 
mass in one loop \cite{19}, but it is also too small.

To obtain phenomenologically interesting radiative neutrino masses, we add 
the $\chi^-$ singlet to produce the following new interactions:
\begin{equation}
f_l (\nu_l \tau_L - l_L \nu_\tau) \chi^+, ~~~ (\phi^+ \eta^0 - \phi^0 \eta^+) 
\chi^- \chi^0,
\end{equation}
where $l = e, \mu$.  The mass-generating radiative mechanism of Ref.~[20] is 
now operative, as shown in Fig.~3.  One should note that the above scalar 
sector contains a pseudo-Goldstone boson which comes about because there are 
3 global U(1) symmetries in the Higgs potential and only 2 local U(1) 
symmetries which get broken.  However, if an extra neutral scalar $\zeta^0$ 
transforming as $(1,1,0;-3)$ is added, then the Higgs potential will have two 
more terms and the extra unwanted U(1) symmetry is eliminated.

An alternative scenario is to replace $\eta$ with a second $\Phi$ doublet, 
but retain $\chi^-$ as well as $\zeta^0$.  In that case, $\langle \phi^0_{1,2} 
\rangle$ break only $SU(2)_L \times U(1)_Y$ whereas $\langle \chi^0 \rangle$ 
and $\langle \zeta^0 \rangle$ break only $U(1)_X$.  In contrast, $\langle 
\eta^0 \rangle$ breaks both.  Hence $X$ has no tree-level flavor-changing 
couplings, and does not mix with $Z$ except through the cross kinetic-energy 
terms \cite{21} which we assume to be negligible.  We show in Fig.~4 the 
one-loop diagram connecting $\nu_\mu$ with $\nu_\tau$.  Note that in this 
scenario, only one linear combination of $\nu_e$ and $\nu_\mu$ picks up a 
nonzero mass in one loop.  The other linear combination will get a mass in 
two loops \cite{18}.  To suppress flavor-changing neutral-current 
interactions in the scalar sector, we impose a discrete $Z_2$ symmetry 
such that $\Phi_1$ is even and $\Phi_2$ is odd so that the latter does not 
couple to leptons.  This discrete symmetry is then broken softly by the 
$\Phi_1^\dagger \Phi_2 + h.c.$ term in the Higgs potential, as in the Minimal 
Supersymmetric Standard Model.

\section{Lepton and Quark Flavor Nonconservation}

In the scenario where we add the scalar $\eta \sim (1,2,1/2;-3)$ doublet, 
the charged-lepton mass matrix linking $\bar e_L$, $\bar \mu_L$, $\bar \tau_L$ 
to $e_R$, $\mu_R$, $\tau_R$ can be chosen to be of the form
\begin{equation}
{\cal M}_l = \left[ \begin{array} {c@{\quad}c@{\quad}c} m_e & 0 & 0 \\ 
0 & m_\mu & 0 \\ a_e & a_\mu & m_\tau \end{array} \right].
\end{equation}
Since $X$ couples only to $\tau$ before symmetry breaking, the fact that 
${\cal M}_l$ is not diagonal induces flavor-changing couplings of $X$ 
to $e$ and $\mu$ as follows:
\begin{equation}
3 g_X X_\nu \left( {a_\mu \over m_\tau} \bar \mu_R \gamma^\nu \tau_R + 
{a_e \over m_\tau} \bar e_R \gamma^\nu \tau_R - {a_\mu a_e \over m_\tau^2} 
\bar e_R \gamma^\nu \mu_R + h.c. \right).
\end{equation}
The best individual bounds on $a_\mu$ and $a_e$ come from the nonobservation 
of $\tau \to \mu \pi^+ \pi^-$ and $\tau \to e \pi^+ \pi^-$.  Assuming that 
the ratios of the above rates to that of $\tau \to \nu_\tau \pi^- \pi^0$ 
are roughly given by those of their inclusive rates, and using the upper 
limits \cite{22} of $7.4 \times 10^{-6}$ and $4.4 \times 10^{-6}$ on their 
branching fractions, we find
\begin{equation}
{g_X^2 \over m_X^2} a_\mu m_\tau < 3.8 \times 10^{-7}, ~~~ {g_X^2 \over m_X^2} 
a_e m_\tau < 2.9 \times 10^{-7}.
\end{equation}
The best bound on the product $a_\mu a_e$ comes from the nonobservation of 
$\mu - e$ conversion in nuclei.  Using the formalism of Ref.~[23] and the 
experimental upper limit of $4.3 \times 10^{-12}$ for Ti, we find
\begin{equation}
{g_X^2 \over m_X^2} a_\mu a_e < 3.1 \times 10^{-12}.
\end{equation}
For $g_X = 0.2$ and $m_X = 60$ GeV, the above bounds translate to
\begin{equation}
a_\mu < 19~{\rm MeV}, ~~~ a_e < 15~{\rm MeV}, ~~~ a_\mu a_e < 0.3~
({\rm MeV})^2.
\end{equation}
We note that\cite{4} a radiative $\nu_\mu$ mass of $2.3 \times 10^{-3}$ eV 
could be obtained with $a_\mu = 10$ MeV.

Independent of possible tree-level lepton flavor nonconservation in any 
variation of the $B - 3 L_\tau$ model, there is quark flavor nonconservation 
in one-loop order involving the $X$ gauge boson.  This contributes to decays 
such as $K^+ \to \pi^+ \nu_\tau \bar \nu_\tau$ and $b \to s \tau^+ \tau^-$. 
However, since $X$ has only vector couplings to quarks, the effective 
one-loop transition amplitude $q_1 \to q_2 X$ has the same form \cite{24} 
as $q_1 \to q_2 \gamma$, {\it i.e.} $k^\mu \epsilon_X^\nu \bar q_2 
\sigma_{\mu \nu} (A + B \gamma_5) q_1$, where $A$ and $B$ are proportional 
to $m_q/M_W^2$.  In contrast, the transition amplitude $q_1 \to q_2 Z$ is 
of the form \cite{25} $\epsilon_Z^\nu \bar q_2 \gamma_\nu (1 - \gamma_5) q_1$. 
Hence the contribution of $X$ to these amplitudes is suppressed by 
$m_q^2/m_X^2$ relative to that of $Z$ and is always negligible.

\section{Conclusion}

Since $B - 3 L_\tau$ can be gauged with the addition of $\nu_{\tau R}$, the 
possible existence of the associated gauge boson $X$ should be investigated. 
We find its coupling $g_X$ and mass $m_X$ to be constrained by the data on 
$Z$ decay in two important ways.  For $m_X < 70$ GeV, the nonobservation of 
the direct decay of $Z$ to $X$ gives an upper limit on $g_X$ as shown in 
Fig.~1.  For $m_X < 50$ GeV, we find a rather stringent limit of $g_X < 0.1$. 
For $m_X > 56$ GeV, a better limit is obtained from the observed $e - \mu - 
\tau$ universality of $Z$ decay as shown also in Fig.~1.  

The $X$ boson may be produced at hadron colliders and be detected through its 
decay into $\tau$ pairs.  The nonobservation of such events at the Tevatron 
puts an upper limit on $g_X$ for $m_X \gsim M_Z$ as shown in Fig.~2.  We 
have also estimated the discovery limits of $X$ at the future Run II of the 
Tevatron and at the LHC as shown also in Fig.~2.  The latter offers a viable
$X$ boson signal upto $m_X \sim 1$ TeV.

Even though only $\nu_\tau$ gets a tree-level mass in the $B - 3 L_\tau$ 
gauge model, the other two neutrinos may also acquire masses and mix with 
$\nu_\tau$ through radiative corrections with an extended scalar sector. 
We have presented two possible scenarios, one with tree-level 
flavor-nondiagonal couplings of the $X$ boson to charged leptons and one 
without.  In the first scenario, important phenomenological constraints 
come from $\tau$ decay and $\mu - e$ conversion in nuclei.  On the other hand, 
quark flavor nonconservation is always suppressed in one loop because 
$X$ couples only vectorially.

In conclusion, if the $X$ boson exists, it may be hiding very effectively 
even if its coupling is not too small and its mass is below $M_Z$. However
the future hadron colliders can probe for the $X$ boson upto a mass range of
$\sim 1$ TeV if its coupling is of the same order of magnitude as $g_1$. 

\vspace{0.3in}
\begin{center}{ACKNOWLEDGEMENT}
\end{center}

We thank J.W. Gary for several useful discussions and 
R. Bhalerao and K. Sridhar for help with the gnuplot program.
This work was supported in part by the U.~S.~Department of Energy
under Grant No.~DE-FG03-94ER40837.

\newpage
\bibliographystyle{unsrt}

\begin{thebibliography}{99}
\bibitem{1} G. 't Hooft, Phys. Rev. Lett. {\bf 37}, 8 (1976); Phys. Rev. 
{\bf D14}, 3422 (1976).
\bibitem{2} X. G. He, G. C. Joshi, H. Lew, and R. R. Volkas, Phys. Rev. 
{\bf D43}, R22 (1991); {\it ibid.} {\bf D44}, 2118 (1991).
\bibitem{3} R. E. Marshak and R. N. Mohapatra, Phys. Lett. {\bf 91B}, 222 
(1980).
\bibitem{4} E. Ma, hep-ph/9709474 (Phys. Lett. {\bf B}, in press).
\bibitem{5} J. C. Pati and A. Salam, Phys. Rev. {\bf D10}, 275 (1974).
\bibitem{6} E. Witten, Phys. Lett. {\bf B117}, 324 (1982).
\bibitem{7} R. Delbourgo and A. Salam, Phys. Lett. {\bf 40B}, 381 (1972); 
T. Eguchi and P. G. O. Freund, Phys. Rev. Lett. {\bf 37}, 1251 (1976); 
L. Alvarez-Gaume and E. Witten, Nucl. Phys. {\bf B234}, 269 (1984).
\bibitem{8} S. L. Adler, Phys. Rev. {\bf 177}, 2426 (1969); J. S. Bell and 
R. Jackiw, Nuovo Cimento {\bf 60A}, 47 (1969); W. A. Bardeen, Phys. Rev. 
{\bf 184}, 1848 (1969).
\bibitem{9} C. D. Carone and H. Murayama, Phys. Rev. Lett. {\bf 74}, 3122 
(1995); Phys. Rev. {\bf D52}, 484 (1995).
\bibitem{10} B. Brahmachari, A. S. Joshipura, S. Rindani, D. P. Roy, and K. 
Sridhar, Phys. Rev. {\bf D48}, 4224 (1993).
\bibitem{11} A. Lopez-Fernandez, J.C. Ramao, F.de Campos and
J.W.F. Valle, Phys. Lett. {\bf B312}, 240 (1993).  ALEPH Collab.,
D. Buskulic et al., Phys. Lett. {\bf B313}, 312 (1993). 
\bibitem{12} ALEPH Collab., D. Buskulic et al., Phys. Lett. {\bf
B384}, 427 (1996).
\bibitem{13} ALEPH Collab., D. Decamp et al., Phys. Rep. {\bf 216},
253 (1992). 
\bibitem{14} ALEPH Collab., D. Buskulic et al., Z. Phys. {\bf C66}, 3
(1995). 
\bibitem{15} The LEP Electroweak Working Group and SLD Heavy Flavour
Group, CERN-PPE/97-154 (Dec 97). 
\bibitem{16} CDF Collab., F. Abe {\it et al.}, Phys. Rev. Lett. {\bf
79}, 3585 (1977); M. Hohlmann (CDF Collab.), Lake Luis Winter School,
Canada (1996).
\bibitem{17} O. Helene, Nucl. Inst. and Methods, {\bf 212}, 319 (1983).
\bibitem{18} K. S. Babu and E. Ma, Phys. Rev. Lett. {\bf 61}, 674 (1988); 
Phys. Lett. {\bf B228}, 508 (1989).  See also S. T. Petcov and S. T. Toshev, 
Phys. Lett. {\bf B143}, 175 (1984).
\bibitem{19} G. C. Branco, W. Grimus, and L. Lavoura, Nucl. Phys. {\bf B312}, 
492 (1989).
\bibitem{20} A. Zee, Phys. Lett. {\bf 93B}, 389 (1980).
\bibitem{21} B. Holdom, Phys. Lett. {\bf B166}, 196 (1986).
\bibitem{22} Particle Data Group, R. M. Barnett {\it et al.}, Phys. Rev. 
{\bf D54}, 1 (1996).
\bibitem{23} J. Bernabeu, E. Nardi, and D. Tommasini, Nucl. Phys. {\bf B409}, 
69 (1993).
\bibitem{24} E. Ma and A. Pramudita, Phys. Rev. {\bf D24}, 1410 (1981).
\bibitem{25} E. Ma and A. Pramudita, Phys. Rev. {\bf D22}, 214 (1980).

\end{thebibliography}


\newpage

\begin{figure}[htbp]
\begin{center}
\leavevmode
\psfig{figure=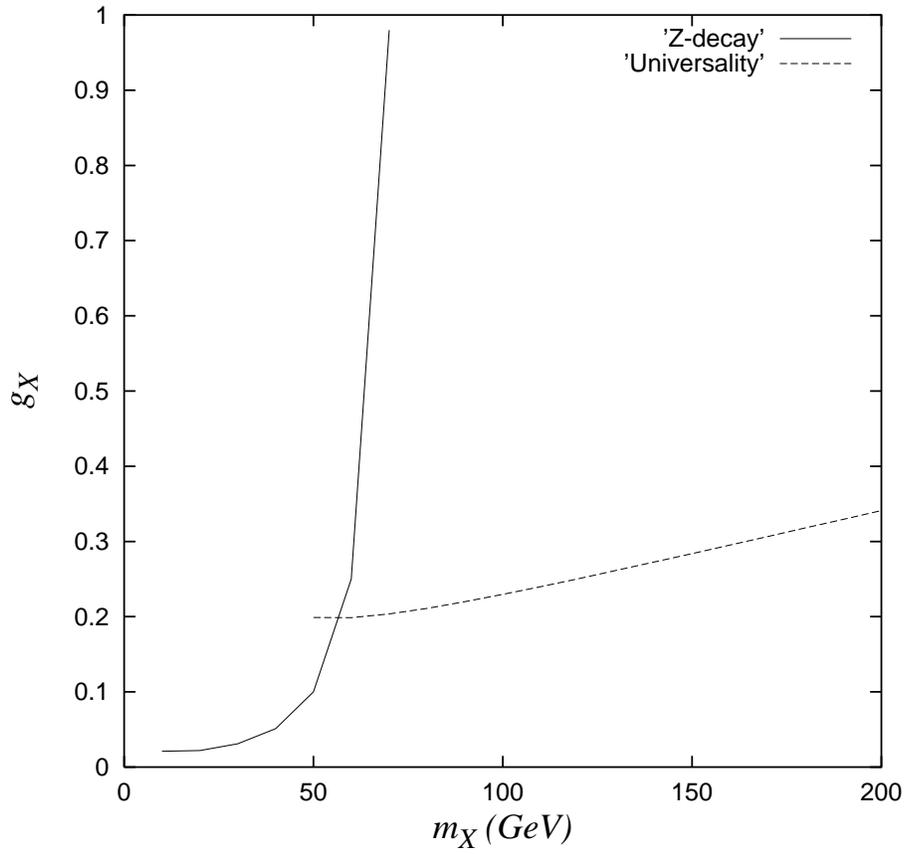,width=12cm}
\caption{The 95\% CL limits on the coupling of the $X$ boson as
functions of its mass coming from $Z$-decay and universality
constraint using LEP-I data.} 
\end{center}
\end{figure}

\newpage

\begin{figure}[htbp]
\begin{center}
\leavevmode
\psfig{figure=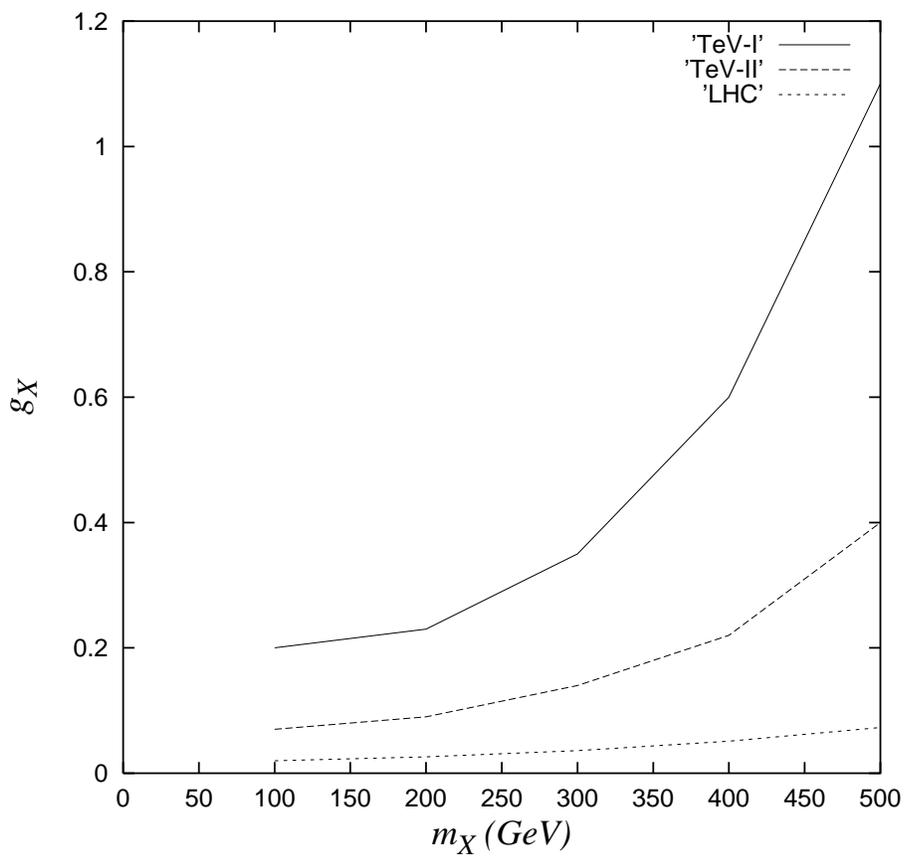,width=12cm}
\caption{The hadron collider limits on the $X$ boson coupling as
functions of its mass.  The present limit from Tevatron (Run I) is
shown along with the discovery limits from Tevatron (Run II) and LHC.} 
\end{center}
\end{figure}

\newpage
\begin{figure}
\begin{center}
\begin{picture}(200,120)(0,0)
\ArrowLine(20,0)(60,0)
\ArrowLine(100,0)(60,0)
\ArrowLine(140,0)(100,0)
\ArrowLine(180,0)(140,0)
\DashArrowLine(68,72)(100,40)6
\DashArrowLine(132,72)(100,40)6
\DashArrowArcn(100,0)(40,180,90)6
\DashArrowArcn(100,0)(40,90,0)6
\DashArrowLine(100,-40)(100,0)6
\Text(40,-8)[c]{$\nu_\mu$}
\Text(80,-8)[c]{$\tau_L$}
\Text(120,-8)[c]{$\mu_R$}
\Text(160,-8)[c]{$\nu_\mu$}
\Text(100,-47)[c]{$\langle \eta^0 \rangle$}
\Text(59,28)[c]{$\chi^-$}
\Text(142,28)[c]{$\phi^-$}
\Text(60,80)[c]{$\langle \chi^0 \rangle$}
\Text(140,80)[c]{$\langle \eta^0 \rangle$}
\end{picture}
\vskip 1.0in
{\bf Fig.~3.} ~Radiative mechanism for neutrino masses in the flavor-changing 
scenario.

\begin{picture}(200,120)(0,0)
\ArrowLine(20,0)(60,0)
\ArrowLine(100,0)(60,0)
\ArrowLine(140,0)(100,0)
\ArrowLine(180,0)(140,0)
\DashArrowLine(100,40)(68,72)6
\DashArrowLine(132,72)(100,40)6
\DashArrowArcn(100,0)(40,180,90)6
\DashArrowArcn(100,0)(40,90,0)6
\DashArrowLine(100,-40)(100,0)6
\Text(40,-8)[c]{$\nu_\mu$}
\Text(80,-8)[c]{$\tau_L$}
\Text(120,-8)[c]{$\tau_R$}
\Text(160,-8)[c]{$\nu_\tau$}
\Text(100,-47)[c]{$\langle \phi^0_1 \rangle$}
\Text(59,28)[c]{$\chi^-$}
\Text(142,28)[c]{$\phi_1^-$}
\Text(60,80)[c]{$\langle \zeta^0 \rangle$}
\Text(140,80)[c]{$\langle \phi_2^0 \rangle$}
\end{picture}
\vskip 1.0in
{\bf Fig.~4.} ~Radiative mechanism for neutrino mass in the flavor-conserving 
scenario.
\end{center}
\end{figure}
\end{document}